**Delta-singular vortex dynamics on a rotating sphere and stability of coupled atmospheric centers of action**

Mokhov I.I. [1], Chefranov S.G. [1], Chefranov A.G. [2,3]

[1] A.M. Obukhov Institute of Atmospheric Physics RAS
(mokhov@ifaran.ru , schefranov@mail.ru)
[2] Taganrog Institute of Technology, South Federal University
[3] Eastern Mediterranean University, North Cyprus
(Alexander.chefranov@emu.edu.tr)

Studies of δ-singular vortex dynamics on a rotating sphere are relatively rare (see [1,2] and references in [2]) though they can be useful for modeling of different hydrodynamic, geophysical and astrophysical systems, in particular for modeling of atmospheric centers of action [3, 4] and blockings [5-7]. It should be noted here a problem of infinite dimension of usually analyzed dynamic systems. Such a problem appears as a result of the invariant absolute vorticity representation (in [1, 2, 6] particularly) on a rotating sphere [8] as a formal superposition of δ-singular point vortices (local vorticity) and background global vorticity (associated with the Earth rotation). Interaction between local and global vorticity (due to the hydrodynamics equations nonlinearity) leads to the structural noninvariance of the local vortices and to the necessity of the approximate considerations.

In this paper, strong interaction of the local and global vorticity due to the introduction of δ-singular (localized) representation of the invariant absolute vorticity is exactly considered. It eliminates the cause of the infinite dimension problem for the appropriate dynamic system. This representation allows to get a new exact weak solution for an ideal incompressible fluid hydrodynamics on a rotating sphere. Such a solution has a form of 2N-dimensional Hamiltonian dynamic system for arbitrary number N≥2 of strongly interacting point diametrically conjugated vortex pairs (CVP) consisting of the point vortexes with the same absolute value of vorticity and different circulation signs that are placed in diametrically conjugated sphere points. It is shown that one such CVP is the only one allowed by hydrodynamic equations elementary $\delta$-singular stationary vortex object on a sphere unlike the sole $\delta$-singular vortex object introduced in [1].

Existence for $\delta$-singular vortices of a stable absolute stationary pair (N=2) vortical cyclonic-anticyclonic solution (not considered earlier) was determined. Such solution for the vortical system in the Northern Hemisphere is possible only in the case with the more intense and more high-latitude cyclonic vortex than the anticyclonic vortex at the same longitude. It agrees with observations for coupled atmospheric centers of action (ACA) in the Northern Hemisphere [3]. Possibility of modeling of the splitting-type blocking [5] with a more intensive anticyclone to the north with respect to the cyclone is also considered. Contrary to the absolute stationary mode of the coupled vortex ACA systems the only relative stationary mode is realized with a finite velocity motion to the west of vortex pair characterizing blocking (with significantly less intensive vortices than for ACA). Significant influence of polar vortices on the stability of the coupled vortex ACA system and vortex pairs simulating blocking is found. Comparison of theoretical results with observed and reanalysis data [3] for coupled quasi-stationary ACA systems over the oceans in the NH is conducted. Exact solution obtained for $N$=2 is used for analysis of approximate results [6] in relation to the vortex modeling problem of atmospheric blockings.

**1.** From three-dimensional hydrodynamic equations for rotating with angular velocity Ω=const ideal incompressible fluid in the absence of radial fluxes with $V_r$=0 it follows absolute vorticity invariance $\omega = \omega_r + 2\Omega\cos\theta$ [8]:

$$\frac{D\omega}{Dt} = \frac{\partial\omega}{\partial t} + \frac{V_\theta}{r}\frac{\partial\omega}{\partial\theta} + \frac{V_\varphi}{r\sin\theta}\frac{\partial\omega}{\partial\varphi} = 0\,. \qquad (1)$$

Here ($r$, $\theta$, $\varphi$) are the spherical coordinates, $V_\varphi = -\frac{1}{r}\frac{\partial\psi}{\partial\theta}, V_\theta = \frac{1}{r\sin\theta}\frac{\partial\psi}{\partial\varphi}$ are components of the non-divergent velocity field, $\psi$ is the stream function related to $\omega_r$, and $\omega_r$ is the radial local vorticity component $\omega_r = -\Delta\psi, \Delta \equiv \frac{1}{r^2\sin\theta}(\frac{\partial}{\partial\theta}\sin\theta\frac{\partial}{\partial\theta} + \frac{1}{\sin\theta}\frac{\partial^2}{\partial\varphi^2})$ is the spherical part of the Laplace operator.

Let find exact weak solution of (1) for the following δ-singular representation of the absolute vorticity field $\omega$ on a sphere of radius $r$=$R$ in the form:

$$\omega = \frac{\Gamma_0}{R^2}(\delta(\theta) - \delta(\theta-\pi)) + \sum_{i=1}^{N}\frac{\Gamma_i}{R^2\sin\theta_i}(\delta(\theta-\theta_i)\delta(\varphi-\varphi_i) - \delta(\theta-\pi+\theta_i)\delta(\varphi-\varphi_i-\pi))\ , \qquad (2)$$

where $\Gamma_i$ are intensities of CVP, characterized by the coordinates $\theta_i, \varphi_i$ , $i=1,..,N$, and $\Gamma_0$ is the intensity of polar vortices.

The stream function is as follows

$$\psi = -\Omega R^2\cos\theta + \frac{\Gamma_0}{2\pi}\ln\frac{1+\cos\theta}{1-\cos\theta} + \frac{1}{2\pi}\sum_{i=1}^{N}\Gamma_i\ln\frac{1+\cos u_i(\theta,\varphi)}{1-\cos u_i(\theta,\varphi)}\ , \qquad (3)$$

where $\cos u_i = \cos\theta\cos\theta_i + \sin\theta\sin\theta_i\cos(\varphi-\varphi_i)$. The first term in the right-hand side of (3) corresponds to the solid body fluid rotation. In (3) it is taken into account that $\psi_0 = \frac{\Gamma_0}{2\pi}\ln\frac{1+\cos\theta}{1-\cos\theta}$ is a solution of the equation $-\frac{1}{\sin\theta}\frac{\partial}{\partial\theta}\sin\theta\frac{\partial\psi_0}{\partial\theta} = \Gamma_0(\delta(\theta) - \delta(\theta-\pi))$. Contrary to the function $\psi_B = \frac{\Gamma}{4\pi} \ln \frac{1}{1-\cos\theta}$ used in [1] (corresponding to the elementary vortex object in [1]), function $\Psi_0$ is an exact stationary solution of the three-dimensional hydrodynamic equations of ideal incompressible fluid, when in the absence of the radial fluxes with $V_r$=0 the radial pressure gradients may be not equal to zero ($\frac{\partial p}{\partial r} \neq 0$), resulting in that $\Psi_B$ does not satisfy to these equations. The function $\Psi_0$ corresponds to the polar system for the point CVPs and after the replacement (from symmetry considerations) of $\cos\theta$ by $\cos u_i$ defines the only elementary δ-singular vortex object on a sphere allowed by the hydrodynamic equations.

It may be shown that (2), (3) correspond to the exact weak solution of the equation (1) when $\theta_i(t)$ and $\varphi_i(t)$ in (2), (3) agree to the 2$N$-dimensional system

$$\frac{d\theta_i}{dt} = \frac{1}{R^2 \sin\theta_i} \frac{\partial\psi(\theta_i,\varphi_i)}{\partial\varphi_i} = -\frac{1}{\pi R^2} \sum_{k=1}^{N} \frac{\Gamma_k \sin\theta_k \sin(\varphi_i - \varphi_k)}{1-\cos^2 u_{ik}},$$

$$\frac{d\varphi_i}{dt} = -\frac{1}{R^2 \sin\theta_i} \frac{\partial\psi(\theta_i,\varphi_i)}{\partial\theta_i} = -\Omega + \frac{\Gamma_0}{\pi R^2 \sin^2\theta_i} - \frac{1}{\pi R^2} \sum_{k=1}^{N} \frac{\Gamma_k (ctg\theta_i \sin\theta_k \cos(\varphi_i - \varphi_k) - \cos\theta_k)}{1-\cos^2 u_{ik}}. \quad (4)$$

Here we use $\Psi$ from (3), $\Gamma_k = const$ for an arbitrary $\Omega$ and $\Gamma_0$, and $\cos u_{ik} \equiv \cos u_i$ for $\theta = \theta_k$ and $\varphi = \varphi_k$. To prove this (as for the vortex dipoles in [9]), it is necessary to substitute (2) in (1), then the result of differentiation to multiply by an arbitrary finite function $\Phi(\theta,\varphi)$ and to integrate it over the sphere surface. After integration by parts (taking into account the non-divergence of velocity field), one gets the system (4) by equating to zero expressions at $\Phi(\theta,\varphi)$ and its same derivatives based on the arbitrariness of $\Phi$.

Obtained in (4) the exact weak solution of the hydrodynamic equations (1) takes into account exactly the sphere rotation effect qualitatively in a different way in comparison to the approximate modeling of the background and local vorticity in [1,2,6], particularly. In other aspects equations (4) (deduced first time herein from hydrodynamic equations) with the accuracy up to the constant factor in the right-hand side of the equation coincide with proposed in [1] from kinematic considerations and widely used (see [10]) equations written for any CVP number N≥2 composed from point vortices. Note that accounting in [11,12] of the sphere rotation effect in the exact solution of the absolute vortex conservation equation (1) practically coincides[1)] with given in (3). An important distinction from [12] is in the use in (3) of the fundamental solution for the spherical part of Laplace operator instead of the operator's eigen functions (as in [12]) which have singularities corresponding

---

[1)] There is correspondence between first members in right-hand side of (3) and in (4) from [12].

to the point vortices on a sphere.

It was suggested in [12] a description excluding these singularities while in (4), vise versa, finite dimensional dynamics of such δ-singular vortices is considered. This difference in the approaches corresponds to the manifestation of the "vortex-wave" dualism for an object with continuous (wave) and discrete (vortex) properties at the same time.

Important distinction of (4) from systems considered in [1,2] is the sign minus at Ω that just corresponds to the exact weak solution (4) of the equation (1) for $\omega$ from (2). Used in (4) stream function form (3) allows exact accounting of the sphere rotation effect in the conservation law (1) and corresponds to the general description based on (4) for time evolution of arbitrary systems of point (logarithmic) vortices on a rotating sphere. It also allows study of stationary vortex modes without restrictions and artificial complications due to conventional (see [13]) stationary reduction of equation (1) leading to the necessity of screened (Bessel) point vortices introduction.

**2.** The system (4) for $N$=2 has relative stationary mode with $\frac{d\theta_i}{dt}=0$ and $\frac{d(\varphi_1-\varphi_2)}{dt}=0$, e.g. in case of equal longitudes $\varphi_1(t)=\varphi_2(t)=\varphi(t)$, if the following equality holds

$$\frac{\Gamma_0(\sin^2\theta_{20}-\sin^2\theta_{10})}{\sin\theta_{10}\sin\theta_{20}}=\frac{\Gamma_1\sin\theta_{10}+\Gamma_2\sin\theta_{20}}{\sin(\theta_{20}-\theta_{10})}, \tag{5}$$

where $\theta_1=\theta_{10}=const, \theta_2=\theta_{20}=const, \theta_{10}\neq\theta_{20}$. It is possible that each vortex moves along the latitudinal circle ($\theta$ is the latitude complement) with constant angular velocity

$$\omega=\frac{d\varphi}{dt}=-\Omega+\frac{\Gamma_0}{2\pi R^2}\left(\frac{1}{\sin^2\theta_{10}}+\frac{1}{\sin^2\theta_{20}}\right)+\frac{\Gamma_1\sin\theta_{10}-\Gamma_2\sin\theta_{20}}{2\pi R^2\sin\theta_{10}\sin\theta_{20}\sin(\theta_{20}-\theta_{10})}. \tag{6}$$

Absolute stationary mode with $\omega$=0 in (6) may exist only when $\Gamma_0\neq0$ or $\Omega\neq0$, and relative with $\omega\neq0$ is possible according to (5) for $\frac{\Gamma_1}{\Gamma_2}=-\frac{\sin\theta_{20}}{\sin\theta_{10}}$ and under $\Gamma_0=0$, $\Omega=0$, when from (6), one gets $\omega=-\frac{\Gamma_2}{\pi R^2\sin\theta_{10}\sin(\theta_{20}-\theta_{10})}$. In particular, for $\Gamma_2<0$ and $\theta_2<\theta_1$ one gets $\omega<0$, that corresponds to the widely spread splitting blocking [5] with an anticyclone (of intensity $\Gamma_2$) to the north of the cyclone (with intensity $\Gamma_1$) and with intensity of the anticyclone greater than that of the cyclone. In that case all the system moves with the angular velocity $|\omega|$ to the west. It is worth to note that in [14] for topographical vortices (in a rotating vessel with the parabolic bottom), it was stated similar relationship for intensities of two vortices and the distance from the axis of rotation. In [13] also very similar ($\frac{\Gamma_1}{\Gamma_2} = -\frac{\sin^2\theta_{20}}{\sin^2\theta_{10}}$ in [13]) condition for the relative stationary mode with $\omega\neq 0$ was obtained for two vortices on a rotating sphere with the use of Bessel vortices and initially stationary reduction of equations (1).

**3.** Condition $\omega=0$ in (6) for $\Omega\neq 0$ and $\Gamma_0=0$ in (5), (6) defines a new stable absolute stationary mode not considered earlier ([13,15], in particular). According to observations [3] it is displayed for coupled ACA above the NH oceans, for instance. It is essential that for this stationary solution, less intensive (compared to the cyclone) anticyclone with $\Gamma_2<0$ (under $\Omega>0$) is always positioned closer to the equator than the cyclone. Actually, according to (5), (6) for $\Gamma_0=0$ and $\omega=0$ with $\theta_{20}>\theta_{10}$ one gets:

$$\Gamma_2 = -\pi R^2\Omega\sin\theta_{10}\sin(\theta_{20}-\theta_{10}). \tag{6a}$$

For the typical ACA positions in the NH one can estimate from (6a) $|\Gamma_2|\approx 10^8\, km^2/day$. It corresponds to the existence of an absolute stationary solution without considering the possible polar vortices effect, in principle allowing decreasing of that estimate to the usually used empirical value $10^6\, km^2/day$ (see [4]). For the stated new absolute stationary solution, there is not only noted qualitative but also quantitative agreement for estimates of the intensities ratio $\Gamma_1/\Gamma_2$ based on observations [3] and theoretical estimates according to the equality $\Gamma_1/\Gamma_2 = -\sin\theta_{20}/\sin\theta_{10}$ under $\theta_{20}>\theta_{10}$ (for $\Gamma_2<0$). For example, in comparison with the data (for 1949-2002, in particular) for the North Atlantic pair Icelandic cyclone – Azores anticyclone, the mean deviation of the theoretical and empirical estimates of the intensities ratio $\Gamma_1/\Gamma_2$, is below 5%. However, respective deviation for the Pacific ocean vortex pair Aleutian cyclone – Hawaiian anticyclone is already remarkably larger (46%).

**4.** Let now $\Gamma_0\neq 0$, $\Omega\neq 0$, and $\sin\theta_{10}\neq\sin\theta_{20}$. In that case, from (5) and (6) for $\omega=0$ negative $\frac{\Gamma_2}{\pi R^2\Omega}<0$ (i.e.. $\Gamma_2$ corresponds to an anticyclone) is possible only under condition $\gamma_1 = \Gamma_1/\Gamma_2 < -\sin\theta_{10}/\sin\theta_{20}$. Then, for example, for $\theta_{20}>\theta_{10}$, it follows from (5) that $\gamma_0 = \Gamma_0/\Gamma_2 > 0$ (i.e. the polar in the NH is also an anticyclone, see Fig. 1) if $\gamma_1 > -\sin\theta_{20}/\sin\theta_{10}$.

Relative stationary solution (5) and absolute stationary solution (5), (6) (for $\omega=0$ in (6)) now (contrary to the case of $\Gamma_0=0$) may be exponentially unstable with respect to the arbitrarily small magnitude disturbances if inequality (A.6) (see Appendix) is violated, corresponding to the case when $R_0 \equiv 0$ in (A.6). Figure 1 gives examples of the use of the pointed out inequality (for $R_0 \equiv 0$ in (A.6)) in comparison to the observed data for ACAs in the Northern Hemisphere [3]. In particular, if the inequality (A.6) (for $R_0 \equiv 0$) is met then the linear stability for $y \approx 25^0$ (Icelandic cyclone) and $z \approx 55^0$ (Azores anticyclone) is provided for negative $\gamma_1$ and $1{,}35<|\gamma_1|<5{,}24$ when according to (5) $\gamma_0 \approx 0{,}3(1{,}94-|\gamma_1|)$.

**5.** In the case of $\sin\theta_{10} = \sin\theta_{20}$ for $\Gamma_0 \neq 0$, particularly for $\theta_{20} = \pi - \theta_{10}$, a more simple inequality $1+4\frac{\Gamma_0}{\Gamma_1}\cos\theta_{10} < 0$ is , instead of (7), a condition of the exponential instability of that stationary solution with $\gamma_1 = -1, \varphi_{10} = \varphi_{20}$.

To compare this condition with the conclusions on the stability in [6] (based on approximate taking into account of the sphere rotation), let's consider an introduced in [6] parameter $G = \frac{\overline{M}_{0z}}{\Gamma_0} = \frac{\Gamma_1}{\Gamma_0}\cos\theta_{10} + \frac{\Gamma_2}{\Gamma_0}\cos\theta_{20} = const$ with $\overline{M}_{0z} = const$ . In particular, $\frac{\Gamma_0}{\Gamma_1} = \frac{2\cos\theta_{10}}{G}$ for $\theta_{20} = \pi - \theta_{10}$ and $\gamma_1 = -1$. For $\theta_{10} < \pi/2$ from the exponential instability condition $1+4\frac{\Gamma_0}{\Gamma_1}\cos\theta_{10} < 0$ it follows that the linear instability takes place only when $G < 0$ and $|G| < G_0$, where $G_0 = 8\cos^2\theta_{10}$. For instance, $G_0$ is about 0.08 for the pair of vortices near equator with $\theta_{10} \approx 84.3^0$ according to [6]. Note that in [6] there was obtained quantitatively close oscillatory instability threshold value with $G_0 = 0.1$. According to (6), the motion of the vortex pair to the west with $\omega<0$ (simulating blocking) may take place for $\Gamma_1 > 0$ when $G < 0$ and $|G| < 4\cos^2\theta_{10}$, and for $\Gamma_1 < 0$ under all $G > 0$ (for $G < 0$ when $|G| > 4\cos^2\theta_{10}$) unlike to [6] (where it is possible only for $G < 0$). Thus, conclusions on the stability of relative stationary mode with $\omega<0$ important for the blocking realization are found to be essentially dependent on the accounting of the finite intensity of polar vortex with $\Gamma_0 \neq 0$. It would be interesting to compare theoretical conditions for the stable blockings possibility obtained here to the results of analysis for polar vortices intensity from observations.

**6.** The exact weak solution of hydrodynamic equations on a sphere in the form (4) allows arbitrariness in the choice of the polar vortices intensity $\Gamma_0$, including also possibility of evolution of $\Gamma_0$ in time $(\Gamma_0(t))$. This arbitrariness may be cancelled if in addition to variability in time for $\Gamma_0$ variability in time for the sphere rotation frequency is allowed $(\Omega = \Omega(t))$. The requirement of preservation of surface integral (for the sphere) of energy $\overline{E}$ (see (A.1) in Appendix) and angular momentum $\overline{M}_\theta$ (see (A.4) taking into account that in the spherical coordinate system $\overline{M}_\varphi = \overline{M}_r = 0$), the system of equations (A.5) for variables $\Omega(t)$ and $\Gamma_0(t)$ can be written out. Accounting of (A.5), in particular, leads to the appearance in the inequality (A.6) of the term $R_0 \neq 0$, which substantially affects on the definition of the stability domains for the stationary mode (5), (6) (when $\omega = 0$ in (6)). Figure 2a shows stability domain (shown by dark-grey) for the stationary mode (5), (6) when $\gamma_1 = -1$ . Figure 2b (illustrating sensitivity of the stability domain to the $\gamma_1$ variation) shows the stability domain when $\gamma_1 = -1{,}001$ . The same time, without taking into account of (A.5), i.e. for $R_0 \equiv 0$ in (A.6), the mode with $\gamma_1 = -1$ is unstable for all $\theta_{10}, \theta_{20}$ when $\sin\theta_{10} \neq \sin\theta_{20}$ .

The case with $\sin\theta_{10} = \sin\theta_{20}$ is considered in section 5 and, in more details, in Appendix.

**7.** In the conducted above analysis of the point CVP dynamics on a rotating sphere, non-uniformity of the subjacent surface, non-adiabatic processes and other factors (including finite size of the vortex core) characteristic for a real atmospheric vortex system were not taken into account. However, it is possible to estimate impact of the purely dynamic component of the vortex interactions on the evolution of large-scale cyclonic and anticyclonic ACA playing an important role in forming of the planetary climate system and its regional peculiarities [3]. Attempts to describe ACA dynamics with the help of the point vortices on a rotating sphere taking into account polar vortexes was undertaken in particular in [4] but without any assessment of these factors contribution in comparison to observations.

Based on the proposed here theoretical approach the appropriate analysis for the vortex modes of ACA pairs from long-term observations [3] was conducted. Analysis of pressure fields allows to estimate parameter $\gamma_1$ (ratio of the circulation intensities for a cyclonic and anticyclonic vortices in the respective ACA pair) used in the theory. For this purpose, data on the pressure anomaly in the each ACA center with respect to its periphery with a given latitude-longitude ACA coordinates (see [3]) were used.

Figure 1 shows examples of the stability domains (indicated by dark-gray color) according to (A.6) (for $R_0 \equiv 0$ ) depending on the complement of the latitude for the cyclonic and anticyclonic vortices of different intensity. Cyclonic and anticyclonic vortices intensity ratios $\gamma_1$ specified on Fig. 1 were realized (see [3]) for the North Atlantic ACAs in winter 1981 (Fig. 1a) and in winter 1988 (Fig. 1b). Respective positions for Icelandic cyclonic and Azores anticyclonic ACAs are noted on Fig. 1 a,b by crosses with the following coordinates: a) $\theta_1$ = 17.5$^o$, $\theta_2$ = 47.5$^o$, b) $\theta_1$ = 30$^o$, $\theta_2$ = 57.5$^o$. According to (A.6) (for $R_0 \equiv 0$) and Fig. 1a the position for the vortex pair of Icelandic and Azores ACAs in winter 1981 corresponds to the instable mode, whereas, in winter 1988 on Fig. 1b, to the stable mode.

To estimate characteristics of variability and instability of the vortex pairs mode from observations the variability of the distance between ACAs along the longitude Δφ for the pairs of Atlantic (Icelandic cyclone and Azores anticyclone) and Pacific (Aleutian cyclone and Hawaiian anticyclone) ACAs was analyzed, in particular. The excess of the standard deviation of Δφ relative to the mean value over the analyzed period (1949-2002) was taken as a criterion of anomalous variability (and instability) for mutual ACAs positions, particularly in winter. Theoretical estimates of the stability/instability of a vortex pair mode correspond to empirical estimates of the longitudinal (Δφ) normality/abnormality with probability 78% for the Icelandic cyclone – Azores anticyclone comply with the empirical estimates of the longitudinal (Δφ) and with remarkably less probability of 54% for the Aleutian cyclone – Hawaiian anticyclone.

One of the factors not included in the proposed above theoretical analysis which must affect on the relative dynamics and stability of the ACA vortex system is the temperature gradient between the oceans and continents. In relation with this, there was conducted analysis of the observed ACA mutual longitudinal positions (Δφ) in the NH depending on the temperature difference ΔT near ocean and land in particular for the winter season according to observations (see http://www.cru.uea.ac.uk./cru/data/). For the North Atlantic ACA pair the probability of concurrent normality and abnormality for Δφ and ΔT was estimated about 67% and about 70% for the North Pacific ACA pair for the period 1949-2002.

According to obtained results contributions of dynamic and thermal factors to the formation of anomalous mutual ACA position are comparable, in winter particularly.

Thus, it was found existence of the stable stationary solution for the Hamiltonian dynamic system of two point vortices of different signs on different latitudes of a rotating sphere in general agreement with observations. It is shown that such a mode realization is possible only in the case when the more intensive cyclonic vortex is on the higher latitude than the anti-cyclonic vortex. Exponential instability criterion for the stationary vortex solution taking into account the polar vortices effect was obtained. Correspondence of the theory with the observed and reanalysis data for the coupled quasi-stationary systems of cyclonic and anticyclonic ACAs in the Northern Hemisphere over oceans was considered.

The authors are grateful for useful discussions to G.S. Golitsyn, V.P. Goncharov, V.M. Gryanik, M.V. Kurganskiy, G.M. Reznik, V.Yu. Tseitlin, O.G. Chhetiani. This study was supported by RFBR and RAS programmes.

## Appendix

1. Dynamic $2N$ - dimensional system (4) provides conservation of integral invariants of kinetic energy $\bar{E}$, angular momentum $\bar{\mathbf{M}}$ and momentum $\bar{\mathbf{P}}$, per mass unit. Here upper bar denotes integration of corresponding values over the sphere surface, and $E = \frac{1}{2}\mathbf{V}^2$, $\mathbf{M} = [\mathbf{r} \times \mathbf{V}]$, $\mathbf{P} = \mathbf{V}$, $\mathbf{V} = \frac{d\mathbf{r}}{dt}$, $\mathbf{r}$ - is a radius-vector in the Cartesian system of coordinates $(x, y, z)$ origin of which is placed in the center of the sphere. Rotation of the sphere with angular velocity $\Omega$ is performed around axis $z$. In the result of averaging one gets $\bar{\mathbf{P}} = 0$, whereas $\bar{E}$ and vector $\bar{\mathbf{M}}$ components are as follows

$$\bar{E} = \frac{1}{8\pi}\sum_{i=1}^{N}\sum_{k=1(i\neq k)}^{N}\frac{\Gamma_i\Gamma_k}{R^2}\ln\frac{1+\cos u_{ik}}{1-\cos u_{ik}} + \\ +\frac{\Gamma_0}{2\pi R^2}\sum_{\theta=1}^{N}\Gamma_k \ln\frac{1+\cos\theta_k}{1-\cos\theta_k} + \frac{4\pi}{3}\Omega^2 R^2 - 2\Omega\left(\Gamma_0 + \sum_{i=1}^{N}\Gamma_i\cos\theta_i\right), \tag{A. 1}$$

$$\bar{M}_z = 2\sum_{i=1}^{N}\Gamma_i\cos\theta_i + 4\Gamma_0 - \frac{8\pi\Omega R^2}{3}, \tag{A. 2}$$

$$\bar{M}_x = 2\sum_{i=1}^{N}\Gamma_i\cos\varphi_i\sin\theta_i, \qquad \bar{M}_y = 2\sum_{i=1}^{N}\Gamma_i\sin\varphi_i\sin\theta_i, \tag{A. 3}$$

where $\theta_i$ and $\varphi_i$, $i = 1, 2 \ldots N$, are the functions of time, defined from solutions of the system (4) with corresponding initial conditions. When $\Omega = 0$ and $\Gamma_0 = 0$ all four values (A.1) – (A.3) are invariants (4), matching with invariants of the dynamic system in [1]. When $\Omega \neq 0$ or $\Gamma_0 \neq 0$ values (A.3) are not invariant, but (A.1) and (A.2) are still invariants of the system (4).

2. Let's note that in the spherical coordinate system for $\mathbf{M}$ and $\mathbf{P}$, we have: $M_r = 0$, $M_\varphi = RV_\theta$, $M_\theta = -RV_\varphi$; $P_r = 0$, $P_\theta = V_\theta$, $P_\varphi = V_\varphi$ (here as before $r = R$, $V_r = 0$). For any $\Omega$ and $\Gamma_0$ there is the only independent non-zero integral value $\bar{M}_\theta$ (because $\bar{P}_\varphi = -\frac{\bar{M}_\theta}{R}$, $\bar{M}_\varphi = 0$, $\bar{P}_\theta = 0$) in the form

$$\bar{M}_\theta = \pi^2 R^2\Omega - 2\pi\Gamma_0 - 2\sum_{i=1}^{N}\Gamma_i\theta_i - \pi\sum_{i=1}^{N}\Gamma_i\,. \tag{A. 4}$$

Value (A.4) is not invariant of system (4) in the general case, since (A. 4) and (A. 2) are not the same.

3. Weak solution (4) allows the dependence of $\Omega$ and $\Gamma_0$ on time. From the conditions $\dot{\bar{E}} = 0$, $\dot{\bar{M}}_\theta = 0$, we get in addition to (4) the system

$$a\dot{\Gamma}_0 \ [\text{illegible}] \ , \qquad 2\dot{\Gamma}_0 - \pi R^2\dot{\Omega} = -\frac{2}{\pi}\sum_{i=1}^{N}\Gamma_i\dot{\theta}_i \ , \tag{A. 5}$$

where $a = \frac{1}{2\pi}\sum_{k=1}^{N}\frac{\Gamma_k}{R^2}\ln\frac{1+\cos\theta_k}{1-\cos\theta_k} - 2\Omega$, $b = -2\bar{M}_{0z} - 2\Gamma_0 + \frac{8\pi R^2\Omega}{3}$, and $\bar{M}_{0z} = \sum_{i=1}^{N}\Gamma_i\cos\theta_i$ is an invariant of the system (4). The use of system (A. 5) together with (4) eliminates arbitrariness in the definition of the value of $\Gamma_0$, allowed when getting a weak solution of equation (1) as a system (4) (since in that case we get only conclusion on constancy of the values $\Gamma_i$ $i = 1, 2...N$ for arbitrary $\Gamma_0$).
When considering stability of the stationary modes (5), (6) (for $\omega = 0$ in (6)) taking into account the system (A.5), it is found out that stability to extremely small disturbances takes place when the following inequality is satisfied

$$\frac{\gamma_1^2 A + \gamma_1 B + C}{\sin^2(z-y)\sin^3 z\sin^2 y} > \frac{4\gamma_0(\cos y\sin^4 z + \gamma_1\sin^4 y\cos z)}{\sin^3 y\sin^4 z} + R_0, \tag{A.6}$$

где

$$z = \theta_{20}, y = \theta_{10}\ (z > y),$$

$$A = \sin^3 y(\sin y - 2\sin z\cos(z-y)), B = -2\sin^2 z\sin^2 y\cos(z-y),$$

$$C = \sin^3 z(\sin z - 2\sin y\cos(z-y))$$

$$R_0 = \frac{4\gamma_1(\sin^2 z - \sin^2 y)(\sin z - \sin y)}{\pi\sin^3 z\sin^2 y(2+R_1)},$$

$$R_1 = \frac{\gamma_1\ln\frac{1+\cos y}{1-\cos y} + \ln\frac{1+\cos z}{1-\cos z} - 2\gamma_2}{4(\frac{2}{3}\gamma_2 - \gamma_0 - \cos z - \gamma_1\cos y)}$$

$$\gamma_0 = \frac{(\gamma_1\sin y + \sin z)\sin y\sin z}{2\sin(z-y)(\sin^2 z - \sin^2 y)},\ \gamma_2 = \frac{\gamma_0(\sin^2 y + \sin^2 z)}{\sin^2 y\sin^2 z} + \frac{\gamma_1\sin y - \sin z}{2\sin(z-y)\sin z\sin y}.$$

In (A.6) as previously $\gamma_1 = \frac{\Gamma_1}{\Gamma_2}, \gamma_0 = \frac{\Gamma_0}{\Gamma_2}$, but $\gamma_2 = 2\pi R^2\Omega/\Gamma_2$.

4. For the stationary mode with с $\theta_{20} = \pi - \theta_{10}$ и $\varphi_{10} = \varphi_{20}$, having place for $\gamma_1 = -1$ only, equations of the small disturbances are as follows:

$$\dot{\theta}_1 = -\frac{\varphi}{4\sin\theta_{10}\cos^2\theta_{10}},$$

$$\dot{\varphi} = \frac{\theta_1(1 - 4\gamma_0\cos\theta_{10})}{\sin^3\theta_{10}}, \tag{A.7}$$

:

where $\varphi = \varphi_1 - \varphi_2$ and $\theta_1$ are extremely small disturbances of the stationary mode $\theta_{10}$, $\theta_{20} = \pi - \theta_{10}$, $\varphi_{10} = \varphi_{20}$, the same time corresponding disturbance $\theta_2$ of the state $\theta_{20}$ is the same as for $\theta_1$ according to the invariant $\bar{M}_{z0}$ (see (A.2)). In (A.7), the value $\gamma_0 = \frac{\pi R^2\Omega}{\Gamma_2}\sin^2\theta_{10} + \frac{1}{4\cos\theta_{10}}$ is defined from the condition of absoluteness of the stationary mode $\theta_{10}$, $\theta_{20} = \pi - \theta_{10}$, $\varphi_{10} = \varphi_{20}$, excluding transport of the vortex pair along the latitude, i.e. for $\omega = 0$ in (6). For (A.7), it follows that absolute stationary mode is stable when $\pi R^2 \Omega s\iota\ ^2\theta_{10} / \Gamma_2 < 0$, i.e. when the vortex with intensity $\Gamma_2$ is anti-cyclonic, and the one with intensity $\Gamma_1$ is cyclonic. It means that namely when accounting finiteness of the intensity of the polar vortices $\gamma_0 \neq 0$, it is possible to have stability of the configuration of two cyclonic vortices placed symmetrically relative to equator. The same time, if to consider not an absolute but a relative stationary mode (with $\omega \neq 0$ in (6)), then for arbitrary $\gamma_0$ with $\gamma_0 < \dfrac{1}{4\cos\theta_{10}}$ the systemсистема (A.7) yields exponentially stable solution for pairs of anti-cyclonic vortices placed symmetrically relative to equator.

For $\sin\theta_{10} = \sin\theta_{20}$, another stationary mode is possible with $\sin\theta_{10} = \sin\theta_{20}$, when $\theta_{10} = \theta_{20}$ and $\varphi_{10} - \varphi_{20} = \pi$. In that case, already with necessity, $\gamma_1 = 1$. Stability of such a stationary state of the pair of identical vortices on the same latitude is defined by a system of equations coinciding with (A.7) but with the change of signs in the write hand of (A.7) on inverse and replacing of $\gamma_0$ by $-\gamma_0$, where now $\gamma_0 = \dfrac{\pi R^2 \Omega}{\Gamma_2} \sin^2\theta_{10} - \dfrac{1}{4\cos\theta_{10}}$. Thus far, stable absolute stationary mode is found to be possible for the case of anti-cyclonic vortices only as for $\gamma_0 = 0$, and for $\gamma_0 \neq 0$ also. The same time, for two identical vortices of any sign, it is possible to have relative stationary mode when the vortices move with the constant speed along the latitude (i.e. for $\omega \neq 0$ in (6)).

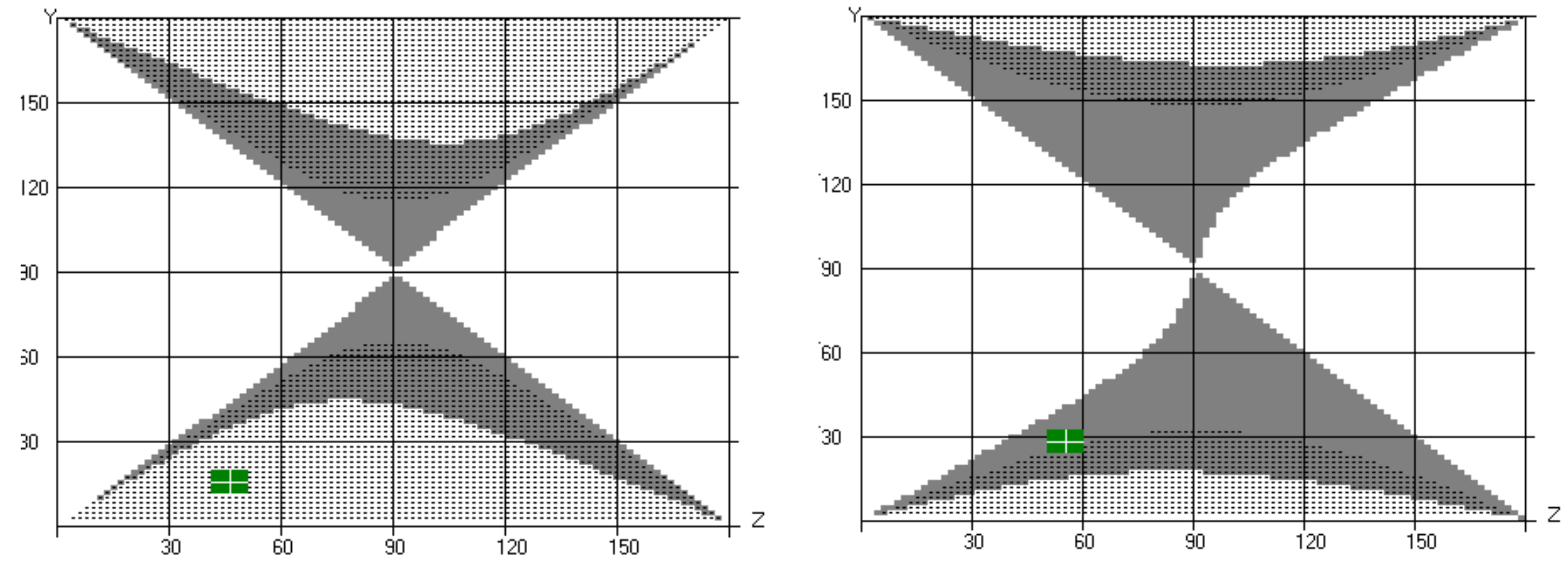

a. $\gamma_1$ = -1.10 b. $\gamma_1$ =-1.87

Fig. 1. Theoretical domains of stability (shown by dark-grey) of the vortex pair on a sphere according to (A.6) (for $R_0 \equiv 0$) depending on the complement to the latitude for anticyclonic (horizontal axis Z) and cyclonic (vertical axis Y) vortices with different values of ratio of their intensities $\gamma_1$ : a) $\gamma_1 = -1.10$ and b) $\gamma_1 = -1.87$ . Observed positions of Icelandic cyclonic and Azores anticyclonic ACAs are shown by crosses: a) for the winter of 1981 with $\gamma_1 = -1.10$; the cross coordinates are $y$ = 17.5$^o$ , $z$ = 47.5$^o$, b) for the winter of 1988 with $\gamma_1 = -1.87$ ; the cross coordinates are $y$ = 30$^o$, $z$ = 57.5$^o$. Shaded domains correspond to positive values of $\gamma_0 = \frac{\Gamma_0}{\Gamma_2}$ - ratio of the polar vortex intensity $\Gamma_0$ to the anticyclonic vortex intensity $\Gamma_2$ .

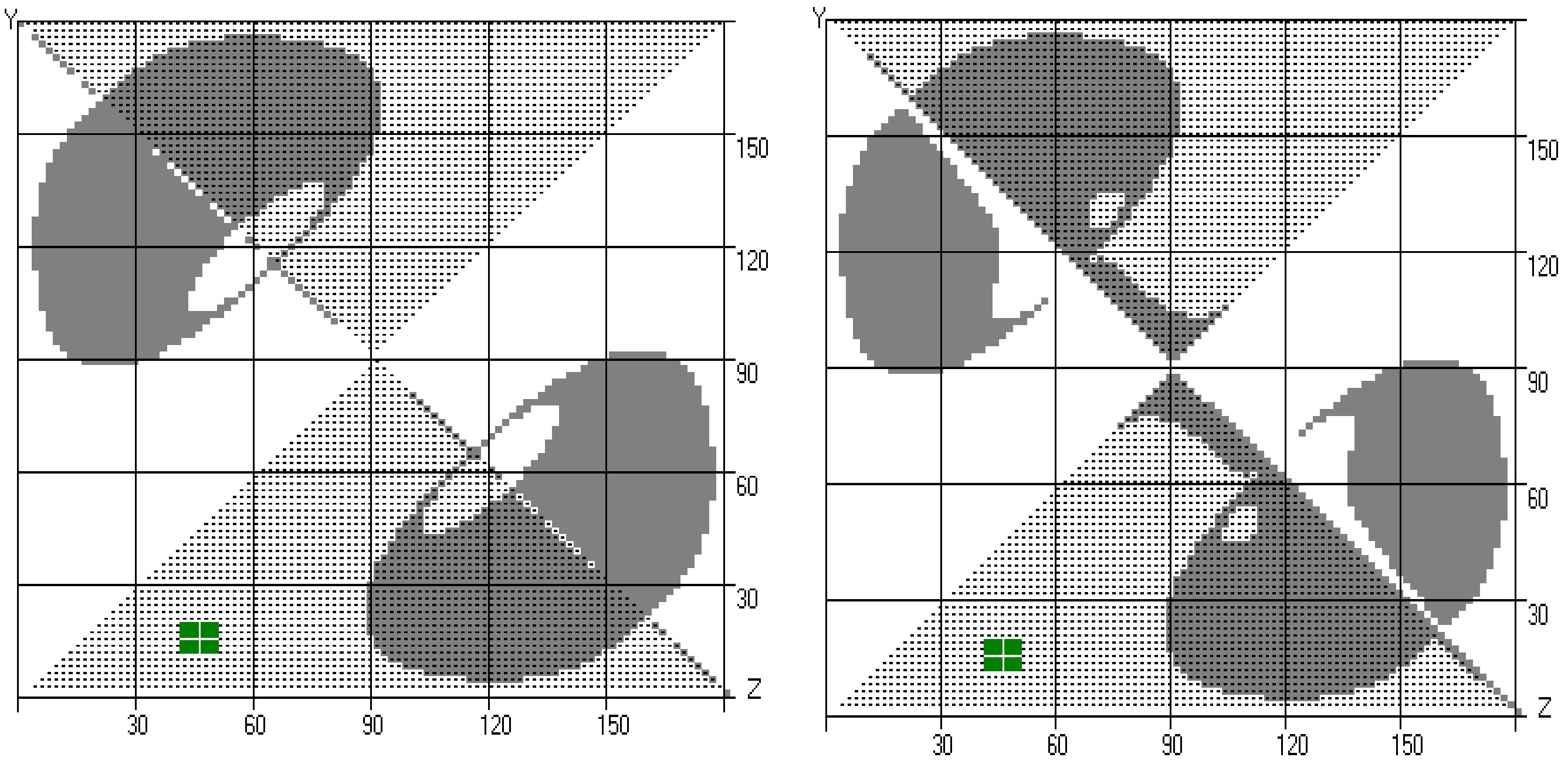

(a) (b)

Fig. 2. Stability domains (shown by dark-grey) for the stationary mode according to (A.6) with $\gamma_1 = -1$ (a) and $\gamma_1 = -1.001$ (b).

**Delta-singular vortex dynamics on a rotating sphere and stability of coupled atmosphere action centers**

I.I. Mokhov, S.G. Chefranov, A.G. Chefranov

Abstract

Existence of a stationary mode for a Hamiltonian dynamic system of two point vortexes with different signs on different latitudes of a uniform rotating sphere complying with observed data is stated. It is shown that such mode realization is possible only in the case when the more intensive cyclonic vortex has greater latitude than that of the anticyclonic vortex. A criterion of exponential instability of the stationary vortex mode taken into account impact of the polar vortexes is obtained. Compliance of the theory to observed data and reanalysis for coupled quasi-stationary systems of cyclonic and anticyclonic atmosphere action centers above oceans in the Northern Hemisphere is considered.